%% file: main.tex
\documentclass[twocolumn,secnumarabic,amssymb, aps, prb, superscriptaddress]{revtex4-2}
\usepackage{graphics, graphicx}
\usepackage[nomath]{stix}
\usepackage[italic]{mathastext}
\usepackage{bm}
\usepackage{bm}
\usepackage{mathtools,amsmath}
\usepackage{adjustbox}
\usepackage{array,booktabs,tabularx}
\usepackage{siunitx}
\usepackage{upquote}
\usepackage{hhline}
\usepackage{stackengine}
\usepackage{xcolor}
\usepackage{setspace} 
\usepackage{hyperref}
\usepackage{microtype}
\usepackage[capitalize]{cleveref}
\usepackage{xr}
\usepackage{float}
\usepackage{xspace}
\usepackage{enumitem}
\usepackage{xcolor}
\hypersetup{
    colorlinks,
    linkcolor={blue!50!black},
    citecolor={blue!50!black},
    urlcolor={blue!80!black}
}

\newcommand\xrowht[2][0]{\addstackgap[.5\dimexpr#2\relax]{\vphantom{#1}}}

\newcommand{\unitu}{cm$^2$/Vs\xspace}

\newcommand{\mue}{$\mu^{\rm e}$\xspace}
\newcommand{\muh}{$\mu^{\rm h}$\xspace}
\newcommand{\frs}{F.R.S.\xspace}
\newcommand{\nkket}{$|n\mathbf{k}\rangle$\xspace}

\newcommand{\Frohlich}{Fr\"ohlich\xspace}

\newcommand{\reH}{$r^{\rm {e,H}}$\xspace}
\newcommand{\rhH}{$r^{\rm {h,H}}$\xspace}

\newcommand{\defomcap}{
  Conduction (a) and valence (c) bands are given in the left column,   the
  corresponding zoomed-in figures (b) and (d) of the shadowed region from
  0 to 2~eV/Bohr are given in the right column.\xspace}

\begin{document}

\title{Guidelines for accurate and efficient calculations of mobilities in two-dimensional materials}

\begin{abstract}
  Emerging two-dimensional (2D) materials bring unprecedented opportunities for
    electronic applications.
  The design of high-performance devices requires an accurate prediction of
    carrier mobility in 2D materials, which can be obtained using state-of-the-art $ab~initio$ calculations.
  However, various factors impact the computational accuracy, leading to
  contradictory estimations for the mobility. In this work,
  targeting accurate and efficient $ab~initio$ calculations, transport properties in III-V monolayers are reported using the Boltzmann
    transport equation, and the influences of pseudopotential,
    quadrupole correction, Berry connection, and spin-orbit coupling (SOC) on
    mobilities are systematically investigated.
  Our findings are as follows: (1)~The inclusion of semi-core states in
    pseudopotentials is important to obtain accurate calculations.
  (2)~The variations induced by dynamical quadrupole and
  Berry connection when treating long range fields can be respectively $40$\% and $10$\%. (3)~The impact of SOC
  can reach up to $100$\% for materials with multi-peak bands.
  Importantly, although SOC notably modifies the electronic wavefunctions, it
    negligibly impacts the dynamical matrices and scattering potential variations.
  As a result, the combination of fully-relativistic electron calculation and
    scalar-relativistic phonon calculation can strike a good balance between
    accuracy and cost.
  This work compares computational methodologies, providing guidelines for
    accurate and efficient calculations of mobilities in 2D semiconductors.
\end{abstract}

\author{Jiaqi Zhou} \email{jiaqi.zhou@uclouvain.be} 
\affiliation{Institute of Condensed Matter and Nanosciences (IMCN), Université catholique de Louvain (UCLouvain), 1348 Louvain-la-Neuve, Belgium}

\author{Samuel Poncé}
\email{samuel.ponce@uclouvain.be}  
\affiliation{Institute of Condensed Matter and Nanosciences (IMCN), Université catholique de Louvain (UCLouvain), 1348 Louvain-la-Neuve, Belgium}
  \affiliation{European Theoretical Spectroscopy Facility}
  \affiliation{WEL Research Institute, Avenue Pasteur 6, 1300 Wavre, Belgium}

\author{Jean-Christophe Charlier} \email{jean-christophe.charlier@uclouvain.be} 
\affiliation{Institute of Condensed Matter and Nanosciences (IMCN), Université catholique de Louvain (UCLouvain), 1348 Louvain-la-Neuve, Belgium}

\date{\today} \maketitle

\section{Introduction}

Two-dimensional (2D) materials exhibit exotic phenomena which can be used in
  electronic and spintronic devices~\cite{Zhang2022Sep,Liu2019Oct,Zhou2023Aug}.
The atomical thickness of 2D semiconductors enables the efficient engineering
  of electronic properties through gate voltage control, making them promising
  for transistor applications~\cite{Wu2022Mar,Shen2022Dec}.
Drift mobility quantifies the transport of carrier in a material in response to
  an electric field, and the prediction of carrier mobility is critical for the
  design of high-performance devices.
A prevalent approach is to employ density functional theory (DFT)
  and density functional perturbation theory (DFPT) to evaluate the
  electron-phonon coupling (EPC) matrices~\cite{Giustino2017Feb}.
However, the accurate calculation of mobility requires very dense momentum
  grids, which is generally too expensive to be performed by direct DFPT calculations.
To solve the problem of computational cost, Wannier functions can be used to
  interpolate EPC from coarse grids to fine grids~\cite{Giustino2007Oct}.
Using this technique, efforts have been devoted to discovering 2D materials with
  high mobilities~\cite{Park2014Mar,Cheng2019Oct,Sohier2018Nov,Sohier2019Jun,
    Ponce2023Apr_PRB,Xiao2023Jul}.
However, conclusions are found to be contradictory due to the various approximations
  used in different studies.
A high-throughput work reports that III-V monolayers with atomically flat
  structures can present high electron mobility (\mue) and high hole mobility
  (\muh).
For example, in the case of BSb monolayer, Refs.~\cite{Zhang2023Feb} and
  \cite{Song2023Apr} report a higher hole mobility (\muh) than electron mobility
  (\mue) with different values: \muh / \mue = 6935 /
  5167~\unitu~\cite{Zhang2023Feb} and \muh / \mue = 16397 /
  9520~\unitu~\cite{Song2023Apr}.
In contrast, Ref.~\cite{Hasani2022Feb} finds a larger \mue~=~16221 than
  \muh~=~7882~\unitu.
These discrepancies illustrate that an accurate prediction for carrier mobility  can be
  complicated since many factors are involved, creating confusion on 
  $ab~initio$ values of computed  mobilities in 2D materials.

The atomic motions in 2D semiconductors generate long-range dynamical dipole and
  quadrupole that should be considered to accurately interpolate EPC.
Moreover, spin-orbit coupling (SOC) can significantly modify the electronic
  structures in materials with heavy elements.
  The semi-core states in pseudopotentials may also influence the hole effective mass.
To understand and facilitate the computation, it is
  necessary to elucidate the impacts on the mobilities in 2D materials caused by
  these factors
  in order to optimize the balance between the accuracy and the computational cost in $ab~initio$ calculations.

In this work, we focus on the III-V monolayers MX (M = Ga, In and X = P, As, Sb)
  considering the moderate bandgap and SOC strength.
The influences on mobilities caused by pseudopotentials with different
  semi-core states~\cite{vanSetten2018May}, dynamical
  quadrupole~\cite{Brunin2020Sep_PRL, Brunin2020Sep_PRB,Jhalani2020Sep}, Berry
  connection~\cite{Ponce2023Apr_PRL,Ponce2023Apr_PRB}, and SOC are systematically
  studied.
Their impacts are interpreted by investigating the momentum- and mode-resolved
  scattering rates.
Besides, DFPT results are found to be negligibly affected by SOC, thus it can be
  neglected to accelerate computation.
After discussing methodologies, the temperature-dependent drift and Hall
  mobilities are computed.
Rather than the unity which is commonly assumed in experiments, it is found
  that the Hall factors range from 1.0 to 1.7.

\setlength{\tabcolsep}{15pt}
\renewcommand{\arraystretch}{1.2}
\begin{table*}[t]
  \centering
  \caption{\label{tab:recipes}
    Methodologies employed in this work.
    PP is the abbreviation for pseudopotential.
    Quad and Berry respectively indicate whether dynamical quadrupole and Berry
      connection are considered.
    FR denotes the fully-relativistic calculation with SOC, while SR denotes the
      scalar-relativistic calculation without SOC.
    $e$ denotes the treatment of electronic  calculation for wavefunctions,
    and $ph$ denotes the treatment of phonon calculation for scattering potential variations.
    The methodology {\#7} strikes a good balance between accuracy and cost
    when a ``mix'' approach is used for the treatment of $e$ and $ph$.
  }
  \begin{tabular*}{0.98\textwidth}{c|c|c|c|c|c|c}
    \hline \hline
    Number & Methodology notation  & PP          & Quad & Berry  & $e$ & $ph$ \\  \hline
    \#1   &valencePP+Quad+Berry+FR   & valence   & Yes  & Yes    & FR  & FR  \\
    \#2  &standardPP+Quad+Berry+FR   & standard  & Yes  & Yes    & FR  & FR  \\
    \#3 &stringentPP+Quad+Berry+FR   & stringent & Yes  & Yes    & FR  & FR  \\
    \#4  &standardPP+Quad+FR         & standard  & Yes  & No     & FR  & FR  \\
    \#5  &standardPP+FR              & standard  & No   & No     & FR  & FR  \\
    \#6  &standardPP+Quad+Berry+SR   & standard  & Yes  & Yes    & SR  & SR  \\
    \textbf{\#7} &\textbf{standardPP+Quad+Berry+mix}  & \textbf{standard}    & \textbf{Yes}  & \textbf{Yes}   & \textbf{FR}  & \textbf{SR}   \\
    \hline \hline
  \end{tabular*}
\end{table*}

\section{Drift mobility}

The phonon-limited drift mobility of carrier in 2D semiconductor is calculated
  as~\cite{Ponce2020Feb,Ponce2018Mar}: \begin{equation}\label{eq:mob}
    \mu_{\alpha\beta} = \frac{-1}{S^{\rm uc}n^{\rm c}}\sum_n \int \frac{\rm{d}^2
    {\mathbf k}}{\Omega^{\rm BZ}} v_{n\mathbf{k}\alpha} \partial_{E_\beta}
    f_{n\mathbf{k}}, \end{equation} where $\alpha$, $\beta$ are Cartesian
  directions and $\alpha=\beta$ in the plane for the III-V monolayers.
The $\partial_{E_{\beta}} f_{n\mathbf{k}} = (\partial f_{n\mathbf{k}}/\partial
  E_\beta)|_{\mathbf{E}=0}$ indicates the linear variation of the electronic
occupation function $f_{n\mathbf{k}}$ in response to the electric field
$\mathbf{E}$.
$S^{\rm uc}$ is the unit cell area, and $\Omega^{\rm BZ}$ the first   Brillouin zone (BZ) area.
The band velocity of state $\varepsilon_{n\bf{k}}$ is given by
  $v_{n\mathbf{k}\alpha} = \hbar^{-1} \partial \varepsilon_{n\mathbf{k}}/\partial
    k_{\alpha}$, and $n^{\rm c}$ denotes the carrier concentration in a vanishing
  limit.
$\partial_{E_{\beta}} f_{n\mathbf{k}}$
can be obtained by solving the linearized Boltzmann transport equation (BTE)~\cite{Ponce2018Mar}:
\begin{equation} \begin{split} \label{eq:BTE} \partial_{E_{\beta}}
    f_{n\mathbf{k}} = &  e v_{n\mathbf{k}\beta} \frac{\partial
      f_{n\mathbf{k}}^0}{\partial \varepsilon_{n\mathbf{k}}} \tau_{n\mathbf{k}} +
    \frac{2\pi\tau_{n\mathbf{k}}}{\hbar} \sum_{m\nu} \int \frac{\rm{d}^2
      \mathbf{q}}{\Omega^{\rm{BZ}}} | g_{mn\nu}(\mathbf{k},\mathbf{q})|^2 \nonumber
    \\ & \times
    \Big[(n_{\mathbf{q}\nu}+1-f_{n\mathbf{k}}^0)\delta(\varepsilon_{n\mathbf{k}} -
    \varepsilon_{m\mathbf{k+q}} + \hbar \omega_{\mathbf{q}\nu} ) \\ & +
    (n_{\mathbf{q} \nu} + f_{n\mathbf{k}}^0)\delta(\varepsilon_{n\mathbf{k}} -
    \varepsilon_{m\mathbf{k+q}} - \hbar \omega_{\mathbf{q}\nu}) \Big]
    \partial_{E_{\beta}} f_{m\mathbf{k}+\mathbf{q}} , \end{split} \end{equation}
where $\tau_{n\mathbf{k}}$ is the total scattering lifetime, and its inverse
$\tau_{n\mathbf{k}}^{-1}$ is the scattering rate, given as \begin{equation}
  \label{eq:scattering_rate} \begin{split} \tau_{n\mathbf{k}}^{-1} & = \frac{2\pi}{\hbar} \sum_{m\nu} \int
    \frac{\rm{d}^2 \mathbf{q}}{\Omega^{\text{BZ}}} | g_{mn\nu}(\mathbf{k,q})|^2 \\
    & \hspace{5mm} \times \big[ (n_{\mathbf{q}\nu} +1 - f_{m\mathbf{k+q}}^0)
      \delta( \varepsilon_{n\mathbf{k}} - \varepsilon_{m\mathbf{k+q}} - \hbar
      \omega_{\mathbf{q}\nu})\\ & \hspace{5mm} + (n_{\mathbf{q}\nu} +
      f_{m\mathbf{k+q}}^0 )\delta(\varepsilon_{n\mathbf{k}} -
      \varepsilon_{m\mathbf{k+q}} + \hbar \omega_{\mathbf{q}\nu}) \big], \end{split}
\end{equation} where $g_{mn\nu}(\mathbf{k,q})$ is the EPC matrix element
denoting the amplitude of scattering between the \nkket state and the
$|m\mathbf{k+q}\rangle$ state via the phonon of frequency
$\omega_{\mathbf{q}\nu}$, $n_{\mathbf{q}\nu}$ is the Bose-Einstein
distribution, and $f_{n\mathbf{k}}^0$ is the Fermi-Dirac occupation function.

\section{Computational details}
To solve these equations, we compute wavefunctions and potential
  variations using the {\sc Quantum ESPRESSO} package~\cite{Giannozzi_2017}.
The norm-conserving pseudopotentials {\sc Pseudo Dojo}~\cite{vanSetten2018May}
  have been employed within the Perdew-Burke-Ernzerhof (PBE) parametrization of
  the generalized gradient approximation (GGA)~\cite{vanSetten2018May}.
Variable cell relaxation is performed with total energy convergence of
  $10^{-8}$~Ry, force convergence of $10^{-4}$~Ry/Bohr, and pressure convergence
  of 0.1~kbar.
The cutoff energy for wavefunctions is set to 120~Ry.
We use the 2D Coulomb truncation scheme of Ref.~\cite{Sohier2017Aug} with a
  vacuum distance over 19~\AA.
Note that in the calculation at the zone-center $\mathbf{q}=\Gamma$, a denser
  $\mathbf{k}$-grid is used in DFPT to get convergent electrostatic properties.
Electron and hole mobilities are calculated using the \textsc{EPW}
  package~\cite{Ponce2016, Lee2023Feb}, where the electron-phonon coupling is
  interpolated from coarse $\mathbf{k}$/$\mathbf{q}$-grids to fine
  $\mathbf{k}$/$\mathbf{q}$-grids using Wannier functions with the considerations
  of dipole, quadrupole, and Berry
  connection~\cite{Ponce2023Apr_PRB,Ponce2023Apr_PRL}.
A coarse $12\times 12\times1$ $\mathbf{k}$/$\mathbf{q}$-grid is adopted in
  $ab~initio$ calculations, then a fine $\mathbf{k}$/$\mathbf{q}$-grid of
  $720\times 720 \times1$ is used for electron mobility, and
  $\mathbf{k}$/$\mathbf{q}$-grid of $360\times 360 \times1$ for hole mobility.
A Fermi surface window of 0.3~eV is used in all calculations.
Convergence tests have been performed with respect to the grid density and
  Fermi surface window.
An adaptive smearing is applied in the energy-conserving delta functions, and a
  phonon frequency cutoff of $1~\rm{cm}^{-1}$ is employed.
The temperature is set at 300~K if not mentioned.
The dynamical quadrupole is computed using linear response as implemented in
  \textsc{Abinit}~\cite{Gonze2020Mar, Royo2019Jun} with a $\mathbf{k}$-grid of
  $32\times 32 \times1$, together with the \textsc{Pseudo Dojo} pseudopotentials
  without non-linear core corrections or spin-orbit coupling~\cite{Dojo}.
For reproducibility, all information including input files, software,
  pseudopotentials, and additional details are provided on the Materials Cloud
  Archive~\cite{MCA}.

\section{Methodology}

In the present section, we focus on the impacts on mobilities induced by
  pseudopotentials, dynamical quadrupole, Berry connection, and SOC.
Using the stringent pseudopotential table from {\sc Pseudo
      Dojo}~\cite{vanSetten2018May} and considering SOC, the atomic structures of all
  the materials have been relaxed.
We find that all the III-V monolayers present a buckled structure, in agreement
  with Refs.~\cite{Hasani2022Feb,Wu2021Jan,Zhuang2013Apr,Sahin2009Oct}.
This buckling breaks the inversion symmetry, leading to a Rashba splitting when
  SOC is considered.
Details related to atomic structures are given in \cref*{supp-sec:struct} in the
Supplemental Material (SM)~\cite{SI2023}.
InAs monolayer is taken as a representative material due to its moderate
  bandgap and SOC strength.
Results for the other monolayers can be found in SM~\cite{SI2023}.
Methodologies are listed in \cref{tab:recipes}.
Three types of control experiments are implemented for the drift mobility
  calculations, respectively evaluating the impacts of pseudopotentials ({\#1},
  {\#2}, and {\#3}), dynamical quadrupole and Berry connection ({\#2}, {\#4}, and
    {\#5}), and fully-relativistic SOC effect ({\#2} and {\#6}).
Another evaluation is performed for the computational efficiency ({\#2} and
    {\#7}) by employing the wavefunctions from the fully-relativistic calculation
  and the scattering potential variation from the scalar-relativistic
  calculation.

\subsection{Impact of pseudopotential}

Three types of norm-conserving pseudopotentials from \textsc{Pseudo Dojo} are
  used in this work.
For elements with the principal quantum number $n\geq3$, the valence
  pseudopotential only includes the valence electrons, the standard
  pseudopotential includes one full $d$ orbital, and the stringent
  pseudopotential includes full $s$, $p$, and $d$ orbitals.
Details of included orbitals can be found in \cref*{supp-tab:pp} in
  SM~\cite{SI2023}.

\begin{figure}[b]
  \includegraphics[width=0.4\paperwidth]{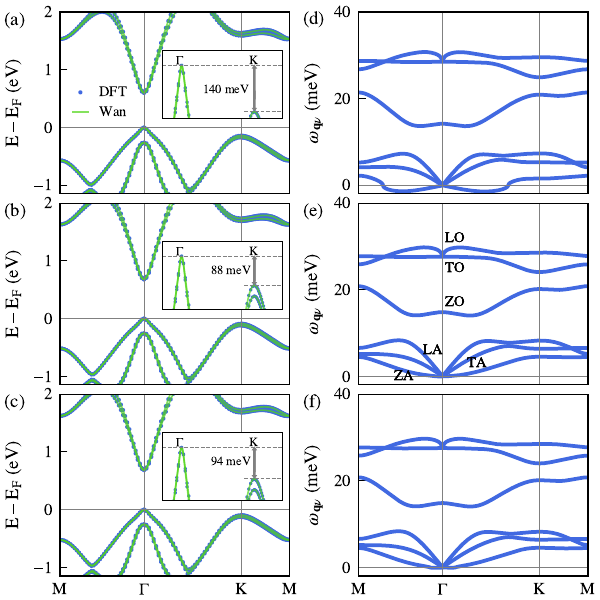}
  \caption{\label{fig:ppbands}
    InAs monolayer electronic band structures (a)-(c) and phonon dispersions (d)-(f) for
      fully-relativistic calculations of valence, standard, and stringent
      pseudopotentials, respectively.
    The insets in (a)-(c) show a zoomed-in figure around the Fermi energy.
    Phonon modes are marked by ZA for the out-of-plane acoustic mode, TA for the
      transverse acoustic mode, LA for the longitudinal acoustic mode, ZO for the
      out-of-plane optical mode, TO for the transverse optical mode, and LO for the
      longitudinal optical mode.
  }
\end{figure}

\Cref{fig:ppbands} shows the electron and phonon dispersions
of InAs monolayer computed by valence, standard, and stringent pseudopotentials, respectively.
Different pseudopotentials introduce small changes to the valence bands.
The valence band maximum (VBM) is always located at the $\Gamma$ point.
For the valence pseudopotential, $\varepsilon_{\rm K} = {\rm E}_{\rm F} -
    140~{\rm meV}$.
However, for the standard and stringent pseudopotentials, the highest valence
  band presents $\varepsilon_{\rm K} \approx {\rm E}_{\rm F} - 90~{\rm meV}$,
  demonstrating that the semi-core states can cause a discrepancy of 50~meV on
  the valence bands.
More intriguingly, phonon dispersions are strongly affected as illustrated in
  Figs.~\ref{fig:ppbands}(d)-(f).
Imaginary frequencies emerge around the $\Gamma$ point in the valence
  pseudopotential calculation.
Besides, the translational invariance is violated by the stringent
  pseudopotential, and the former can be recovered by the acoustic sum rule as
  demonstrated by \cref*{supp-fig:InAsB} in SM~\cite{SI2023}.
Overall, the quadratic ZA mode is derived with the standard
  pseudopotential.

\begin{figure}[b]
  \includegraphics[width=0.4\paperwidth]{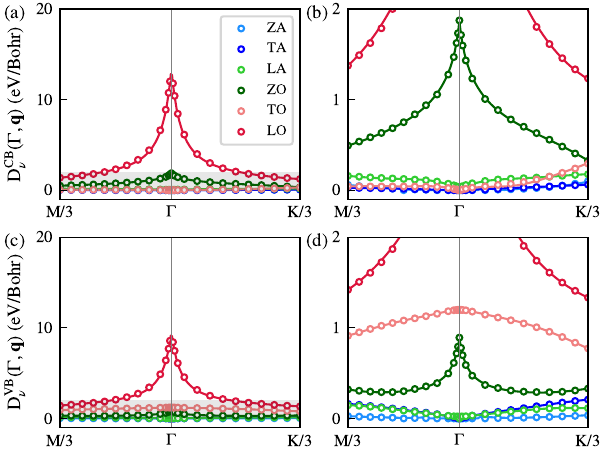}
  \caption{\label{fig:ppdef}
    Comparison between the deformation potentials of InAs monolayer with the
      initial state ${\mathbf k}=\Gamma$ along high-symmetry lines, the DFPT and
      Wannier interpolation results are respectively denoted by dots and lines.
    The standard pseudopotential is employed.
    \defomcap
  }
\end{figure}

\setlength{\tabcolsep}{5pt}
\renewcommand{\arraystretch}{1.2}
\begin{table*}[t]
  \centering
  \caption{\label{tab:electrostatic}
    Dynamical dipole $Z$ ($e$) and quadrupole $Q$ ($e$Bohr), separation length $L$
      (Bohr), dielectric and polarizability tensors ($\varepsilon$ and $\alpha$) of
      all the pristine monolayers.
    Only independent components are presented.
    Note that the crystal symmetry constants $Z_{\kappa xx} = Z_{\kappa yy}$,
      $Q_{\kappa xxz} = Q_{\kappa yyz}$, $Q_{\kappa xxy} = Q_{\kappa yxx} =
        -Q_{\kappa yyy}$, $Q_{\kappa zxx} = Q_{\kappa zyy}$, $\varepsilon^{||}_{xx} =
        \varepsilon^{||}_{yy}$, and $\alpha^{||}_{xx} = \alpha^{||}_{yy}$.
  }  
  \begin{tabular*}{1\textwidth}{c|cc|cc|cc|cc|cc|cc} 
    \hline \hline \xrowht[()]{6pt}
    2DM & Ga & P & In & P & Ga & As & In & As & Ga & Sb & In & Sb \\ \hline
    $Z_{\kappa xx}$ & 2.956 &-2.956 & 3.223 &-3.223 & 2.946 &-2.946 & 3.222 &-3.222 & 2.765 &-2.765 & 3.039 &-3.039 \\
    $Z_{\kappa zz}$ & 0.132 &-0.132 & 0.153 &-0.153 & 0.089 &-0.089 & 0.118 &-0.118 & 0.046 &-0.046 & 0.075 &-0.075 \\
    $Q_{\kappa xxz}$ & -0.858 &-1.209 & -1.007 &-1.809 & -1.444 &-1.650 & -0.260 &-5.396 & -1.996 & -1.578 & -2.139 & -2.110 \\
    $Q_{\kappa xxy}$ & 14.158 &-4.579 & 15.039 &-4.345 & 16.900 &-7.106 & 18.077 &-6.348 & 21.026 &-10.996 & 21.485 &-10.062 \\
    $Q_{\kappa zxx}$ &-11.905 &7.813 &-17.423 &11.754 &-26.850 &20.669 &-34.642 &26.327 &-41.400 & 34.332  &-45.725 & 37.248 \\
    $Q_{\kappa zzz}$ & 0.638 &-0.569 & 0.775 &-0.666 & 1.150 &-1.070 & 2.599 &-6.637 & 1.454 & -1.352 & 1.553 & -1.434 \\
    $L$ &\multicolumn{2}{c|}{9.852} &\multicolumn{2}{c|}{12.257} &\multicolumn{2}{c|}{16.218} &\multicolumn{2}{c|}{18.973} &\multicolumn{2}{c|}{27.624} &\multicolumn{2}{c}{26.270} \\
    $\varepsilon^{||}_{xx}$ &\multicolumn{2}{c|}{3.034} &\multicolumn{2}{c|}{3.408} &\multicolumn{2}{c|}{3.874} &\multicolumn{2}{c|}{4.449}
    &\multicolumn{2}{c|}{5.372} &\multicolumn{2}{c}{5.768} \\
    $\varepsilon^{\perp}$ &\multicolumn{2}{c|}{1.152} &\multicolumn{2}{c|}{1.160} &\multicolumn{2}{c|}{1.161} &\multicolumn{2}{c|}{1.169}
    &\multicolumn{2}{c|}{1.174} &\multicolumn{2}{c}{1.180} \\
    $\alpha^{||}_{xx}$ &\multicolumn{2}{c|}{6.117} &\multicolumn{2}{c|}{7.245} &\multicolumn{2}{c|}{8.644} &\multicolumn{2}{c|}{10.375}
    &\multicolumn{2}{c|}{13.150} &\multicolumn{2}{c}{14.341} \\
    $\alpha^{\perp}$ &\multicolumn{2}{c|}{0.458} &\multicolumn{2}{c|}{0.483} &\multicolumn{2}{c|}{0.486} &\multicolumn{2}{c|}{0.509}
    &\multicolumn{2}{c|}{0.525} &\multicolumn{2}{c}{0.544} \\ \hline \hline
  \end{tabular*} \end{table*}

The convergence of mobility requires a fine grid to sample BZ, making DFPT
  calculation too expensive to be afforded.
Thus, maximally localized Wannier functions (MLWFs) are used for EPC
  calculation.
To assess the quality of the Wannier interpolation, we compare the interpolated
  band structures in Figs.~\ref{fig:ppbands}(a)-(c), and EPC matrix element with
  those obtained from a direct DFPT calculation.
For simplicity, we compute the total deformation potential~\cite{Ponce2021Oct}
  in the BZ center zone: \begin{equation}\label{eq:deformation} D_{\nu}(\Gamma,
  \mathbf{q}) = \frac{1}{\hbar N^{\rm w}}\bigg[ 2 \rho \Omega^{\rm}
  \hbar\omega_{\mathbf{q}\nu} \!
\sum_{nm}  |g_{mn\nu}(\Gamma,\mathbf{q})|^2  \bigg]^{1/2},
\end{equation}
where the $\mathbf{k}=\Gamma$ point is chosen, the sum over bands is carried over the $N^{\rm w}$ states of the Wannier manifold, and $\rho$ is the mass density of the crystal.
It is found that deformation potentials calculated by different
  pseudopotentials are quite similar, thus only the result of standard
  pseudopotential is presented in \cref{fig:ppdef}, results of valence and
  stringent pseudopotentials are given in \cref*{supp-fig:DefPP} in
  SM~\cite{SI2023}.
The Wannier interpolation reproduces the direct DFPT calculation quite well,
hence 
  validating its quality as well as the following
  computed transport properties.
It should be noted that the convergence of DFPT calculations of
  $\mathbf{q}$-points around $\Gamma$ requires a denser $\mathbf{k}$-grid of DFT
  calculations.
Here a $32\times32\times1$ $\mathbf{k}$-grid is employed at ${\bf q} = \Gamma$,
  a $18\times18\times1$ $\mathbf{k}$-grid is used for $\mathbf{q}$-points around
  $\Gamma$, and a $12\times12\times1$ $\mathbf{k}$-grid is employed for
  $\mathbf{q}$-points around M/3 and K/3.

\begin{figure}[b]
  \includegraphics[width=0.4\paperwidth]{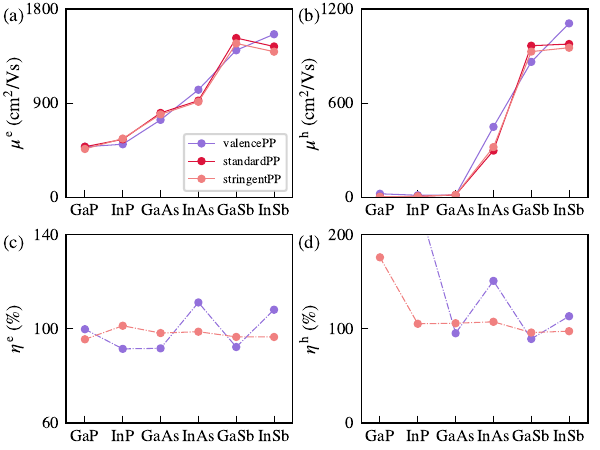}
  \caption{\label{fig:ppmob}
  Electron (a) and hole (b) mobilities calculated using different
pseudopotentials according to \#1, \#2, and \#3 methodologies. 
Taking the result of standard pseudopotential as a benchmark, variations of the
results of valence and stringent pseudopotentials are given as ratios in
(c) for electron and (d) for hole, respectively. 
  }
\end{figure}

\Cref{fig:ppmob}
displays the mobilities of the III-V monolayers.
Large electron mobilities around 1000~\unitu are presented, while the hole
  mobilities of GaP, InP, and GaAs are significantly suppressed, and relatively
  large $\mu^{\rm h}$ are found in InAs, GaSb, and InSb.
To have a more intuitive comparison, the ratios of the valence and stringent
  pseudopotential results to the standard pseudopotential result are displayed in
  Figs.~\ref{fig:ppmob}(c) and (d), respectively for electron and hole
  mobilities.
For the electron mobility, the variation is limited to $\sim10\%$.
For the hole mobility, considerable variations are observed: InAs monolayer
  presents $\eta^{\rm h}=151\%$, and two phosphides even present $\eta^{\rm h}$
  over 200\%.
It should be noted that in the valence pseudopotential results, imaginary
  frequencies appear in the phonon dispersions of InAs, GaP, and InP monolayers as
  shown in \cref{fig:ppbands}(d), \cref*{supp-fig:GaPB}, and \cref*{supp-fig:InPB}
  in SM~\cite{SI2023}, thus the calculated mobility may be questioned.
Besides, the standard and stringent pseudopotentials present consistent
  behaviors in most cases, except for the $\mu^{\rm h}$ of GaP since it is
  minimal thus vulnerable to numerical errors for a ratio.
The above discussion shows that the semi-core states will influence the
  electronic structure and phonon dispersions, leading to variations in
  mobilities.
Considering that standard and stringent pseudopotentials present consistent
  behaviors, applying the former can produce correct phonon dispersion without
  acoustic sum rule, we conclude that the standard pseudopotential is the most
  appropriate for mobility calculation.

\subsection{Impacts of dynamical quadrupole and Berry connection}
In the case of infrared active 2D materials, the \Frohlich interaction leads to a direction-dependent $g_{mn\nu}(\mathbf{k},\mathbf{q})$ for $|{\bf q}|~\to~0$~\cite{Vogl1976Jan}, hindering the application of Wannier interpolation.
This non-analytic behavior can be solved by separating
  $g_{mn\nu}(\mathbf{k},\mathbf{q})$ into a short-range ($\mathcal{S}$) and long-range
  ($\mathcal{L}$) contribution~\cite{Verdi2015Oct}, as

  \begin{equation}\label{geqfirst} g_{mn\nu}(\mathbf{k},\mathbf{q}) =
    g_{mn\nu}^{\mathcal{S}}(\mathbf{k},\mathbf{q}) +
    g_{mn\nu}^{\mathcal{L}}(\mathbf{k},\mathbf{q}), \end{equation}

  \noindent where the short-range part is smooth and analytic
  in $\mathbf{q}$.
The non-analyticity is included in the long-range part, which is given by an
  explicit expression.
Once it is obtained, $g_{mn\nu}(\mathbf{k},\mathbf{q})$ can be accurately calculated: firstly
  subtracting $g_{mn\nu}^{\mathcal{L}}(\mathbf{k},\mathbf{q})$ to obtain $g_{mn\nu}^{\mathcal{S}}(\mathbf{k},\mathbf{q})$ on
  the coarse grid, then applying the Wannier interpolation to
  $g_{mn\nu}^{\mathcal{S}}(\mathbf{k},\mathbf{q})$, and finally adding back $g_{mn\nu}^{\mathcal{L}}(\mathbf{k},\mathbf{q})$ to
  the interpolated $g_{mn\nu}^{\mathcal{S}}(\mathbf{k},\mathbf{q})$ on the fine grid to obtain a
  complete $g_{mn\nu}(\mathbf{k},\mathbf{q})$.
Using the long-range scattering potential
  $V_{\mathbf{q}\kappa\alpha}^{\mathcal{L}}$ which refers to the displacement of
  atom $\kappa$ in the direction $\alpha$ along a phonon mode $\mathbf{q}$, the expression of
  $g_{mn\nu}^{\mathcal{L}}(\mathbf{k},\mathbf{q})$ is given as~\cite{Verdi2015Oct,Ponce2023Apr_PRL}

  \begin{multline}\label{eq:multipole}
    g^\mathcal{L}_{mn\nu}(\mathbf{k,q}) = \Big[ \frac{\hbar}{2
        \omega_{\nu}(\mathbf{q})}\Big]^{\frac{1}{2}} \sum_{\kappa\alpha}
    \frac{e_{\kappa\alpha\nu}(\mathbf{q})}{\sqrt{M_{\kappa}}} \sum_{sp}\\ \times
    U_{ms\mathbf{k+q}} \langle u_{s\mathbf{k+q}}^{\rm W}|
    V_{\mathbf{q}\kappa\alpha}^{\mathcal{L}} | u_{p\mathbf{k}}^{\rm W} \rangle
    U_{pn\mathbf{k}}^{\dagger}, \end{multline}

  \noindent where
  $e_{\kappa\alpha\nu}$ is the phonon eigenvector, $M_{\kappa}$ is the atomic
  mass, $U_{ms\mathbf{k}}$ denotes the Wannier rotation matrices applied to the
  periodic part of the wavefunctions expressed in the Wannier basis $|
    u_{n\mathbf{k}}^{\rm W} \rangle$.

Truncating the expansion at the order of $\mathcal{O}(\mathbf{q})$, the
  long-range scattering potential is given as~\cite{Ponce2023Apr_PRB}
  \begin{multline}\label{eq:vlr_lw}
    V_{\mathbf{q}\kappa\alpha}^{\mathcal{L}}(\mathbf{r}) = \frac{\pi e}{S^{\rm uc}}
    \frac{f(|\mathbf{q}|)}{|\mathbf{q}|} e^{-i \mathbf{q} \cdot
        \boldsymbol{\tau}_\kappa} \bigg[
    \frac{1}{\tilde{\epsilon}^\parallel(\mathbf{q})} \Big\{ 2 i\mathbf{q} \cdot
    \mathbf{Z}_{\kappa\alpha} \\ + \mathbf{q}\cdot \mathbf{q} \cdot
    \mathbf{Q}_{\kappa\alpha} - |\mathbf{q}|^2Q_{\kappa\alpha zz} - 2 \mathbf{q}
    \cdot \mathbf{Z}_{\kappa\alpha} \mathbf{q} \cdot
    V^{\textrm{Hxc},\boldsymbol{\mathcal{E}}}(\mathbf{r})/e \Big \} \\ +
    \frac{1}{\tilde{\epsilon}^\perp(\mathbf{q})} \Big\{ 2|\mathbf{q}|^2
    Z_{\kappa\alpha z} \big[ z + V^{{\rm Hxc},\mathcal{E}_z} (\mathbf{r})/e
      \big]\Big\} \bigg], \end{multline} where 
      $V^{\textrm{Hxc},\boldsymbol{\mathcal{E}}}$
denotes  the self-consistent potential change
induced by the electric field perturbation, $\boldsymbol{\tau}_{\kappa}$ denotes
  the position of atom $\kappa$ within the cell, $\mathbf{Z}_{\kappa\alpha}$ is
  the dynamical dipole along $\alpha$ direction, and $\mathbf{Q}_{\kappa\alpha}$
  is the dynamical quadrupole.
The range separation function $f(|\mathbf{q}|) = 1 - \tanh(|\mathbf{q}| L / 2)$
  is a low-pass Fourier filter that ensures the macroscopic character of the
  potential, where the parameter $L$ defines the length scale.
$\tilde{\epsilon}^\parallel $
and $\tilde{\epsilon}^\perp$ are the dielectric functions given by Eqs.~(38)-(39) in Ref.~\cite{Ponce2023Apr_PRB}.
The above parameters of III-V monolayers are given in \cref{tab:electrostatic}.

Beside the expansion of $V_{\mathbf{q}\kappa\alpha}^{\mathcal{L}}$, the
  Wannier-gauge eigenstates are also expanded to the first order of Taylor series
  as

  \begin{equation}\label{eq:uexp} \langle
  u_{s\mathbf{k+q}}^{\rm W}| = \langle u_{s\mathbf{k}}^{\rm W}| + \sum_\alpha
  q_\alpha \left\langle \frac{\partial u_{s\mathbf{k}}^{\rm W}}{\partial
    k_\alpha}\right|.
\end{equation}

\begin{figure}[t]
  \includegraphics[width=0.4\paperwidth]{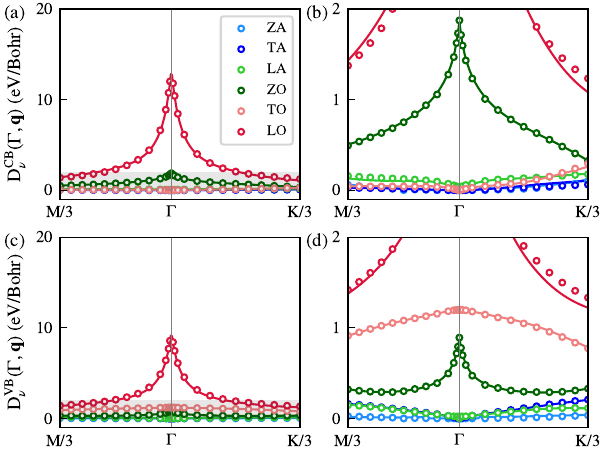}
  \caption{\label{fig:Quad_noBerryD}
    Comparison between the InAs deformation potentials of DFPT and Wannier
      interpolation.
    Dynamical quadrupoles are considered while the Berry connection is neglected
      for the Wannier interpolation.
    \defomcap
  }
\end{figure}
\begin{figure}[t]
  \includegraphics[width=0.4\paperwidth]{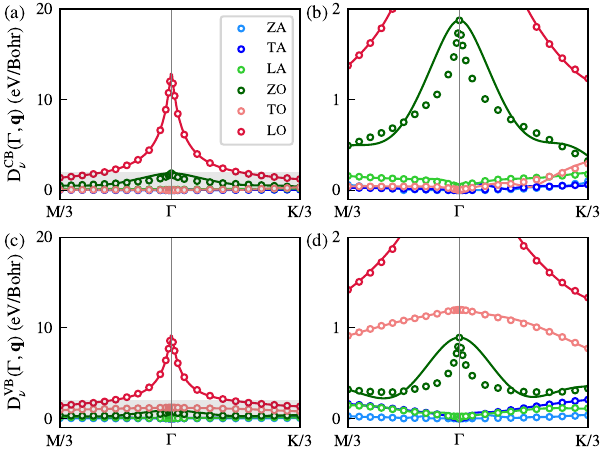}
  \caption{\label{fig:noQuad_noBerryD}
    Comparison between the InAs deformation potentials of DFPT and Wannier
      interpolation in the absence of both dynamical quadrupole and Berry connection.
    \defomcap
  }
\end{figure}

\noindent By
introducing the Berry connection $A_{sp\mathbf{k},\alpha}^{\text{W}} \equiv -i
  \langle \frac{\partial u_{s\mathbf{k}}^{\rm W}}{\partial k_\alpha} |
  u_{p\mathbf{k}}^{\rm W} \rangle$
and ${\varphi}_{\mathbf{q}}$ as the form factor
for $V_{\mathbf{q}\kappa\alpha}^{\mathcal{L}}$,
we obtain

\begin{equation}\label{eq:uexp}
  \langle u_{s\mathbf{k+q}}^{\rm W}| {\varphi}_{\mathbf{q}}  |
  u_{p\mathbf{k}}^{\rm W} \rangle \approx \delta_{sp} \!
  +\! i \mathbf{q} \! \cdot \!
  \Big[  \langle u_{s\mathbf{k}}^{\rm W}| V^{\boldsymbol{\mathcal{E}}}| u_{p\mathbf{k}}^{\rm W} \rangle  \!+\!  \mathbf{A}_{sp\mathbf{k}}^{\text{W}} \Big].
\end{equation}
The term of $V^{\boldsymbol{\mathcal{E}}}$ in \cref{eq:uexp} is omitted in our
  calculations due to its negligible
  contribution~\cite{Brunin2020Sep_PRL,Ponce2023Apr_PRL}.
The term of $\mathbf{A}_{sp\mathbf{k}}^{\text{W}}$ involves the Berry
  connection, which is crucial since it ensures the smoothness of
  $g_{mn\nu}^{\mathcal{S}}(\mathbf{k},\mathbf{q})$ and restores gauge covariance to the lowest order in
  \textbf{q}~\cite{Ponce2023Apr_PRL,Ponce2023Apr_PRB}.

\begin{figure}[b]
  \includegraphics[width=0.4\paperwidth]{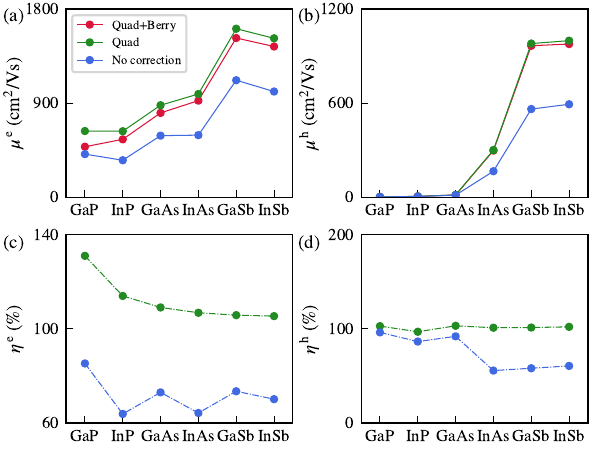}
  \caption{\label{fig:quadmob}
  Electron (a) and hole (b) mobilities calculated with and
  without corrections according to \#2, \#4, and \#5 methodologies. 
  Taking the result with quadrupole and Berry connection as a benchmark,
  variations of the results of neglecting Berry connection and neglecting
  both corrections are given as ratios in (c) for electron and (d) for hole,
  respectively. 
  }
\end{figure}

The effects of two corrections mentioned above, namely the
  dynamical quadrupole and the Berry connection, on the carrier
  mobility in III-V 2D semiconductors are investigated in the
  present section. 
The standard pseudopotential and SOC are applied to the calculations.
The quality of the Wannier interpolations are checked with the deformation potentials
  of the InAs monolayer.
\Cref{fig:Quad_noBerryD} shows the result of the Wannier interpolation
without Berry connection,
which leads to slight deviations from DFPT results for the LO modes.
\Cref{fig:noQuad_noBerryD} presents
the  Wannier result in the absence of both Berry connection and dynamical quadrupole.
For the ZO mode, neglecting quadrupole introduces large deviations from the
  DFPT results for both conduction and valence bands. 

\begin{figure}[t]
  \includegraphics[width=0.38\paperwidth]{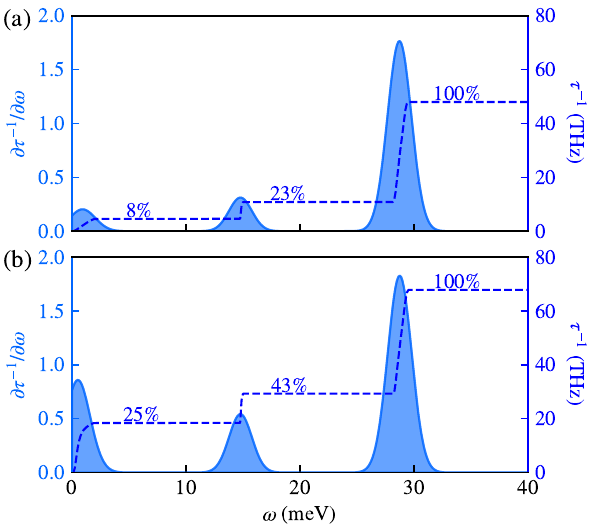}
  \caption{\label{fig:quadfreq}
    Spectral decomposition of the electron scattering rates as a function of phonon
      energy of InAs monolayer (a)~in the presence and (b)~in the absence of both dynamical
      quadrupole and Berry connection.
    The peaks represent $\partial \tau ^{-1} / \partial \omega$ (left axis), and
      the dashed lines represent the cumulative integral $\tau ^{-1}$ (right axis).
    The percentages indicate the cumulative contribution to the total value of $\tau ^{-1}$.
  }
\end{figure}

\Cref{fig:quadmob} shows
the impacts on the mobilities
caused by the two types of corrections.
For \mue in phosphides, the impact of Berry connection is large since the ZA,
  ZO, and TO modes are not well interpolated, see details in \cref*{supp-fig:GaPD}
  and \cref*{supp-fig:InPD} in SM~\cite{SI2023}.
In other cases, the impact of the Berry connection is limited to 10\%,
  attributed to the relatively good fittings as manifested by the InAs result in
  \cref{fig:Quad_noBerryD}.
Further neglecting dynamical quadrupole leads to a mobility underestimation for
  both electron and hole for all the III-V monolayers.
The underestimation can reach up to 40\% for the electron mobilities of all the
  materials, as well as for the hole mobilities in InAs and antimonides.
Regarding phosphides and GaAs where the hole mobilities are strongly suppressed
  by band structures with multi-peaks in BZ, the effect of quadrupole is also
  suppressed.
Note that this conclusion, i.e. neglecting quadrupole will underestimate the
  mobility, can not be generalized for all the materials, since scattering
  rates depend on not only the EPC matrix elements but also on the conservation of
  energies and momenta. 
Moreover, \cref{fig:noQuad_noBerryD} shows that the behavior of Wannier
  deformation potential of ZO mode is not systematic, namely, Wannier deformation
  potential is higher than the DFPT around $\Gamma$ but lower away from $\Gamma$.
Thus, the influence of neglecting quadrupole can only be given by calculations.
In any case, neglecting quadrupole will lower the quality of the Wannier
  interpolation, leading to unconvincing computational results.

The behavior of the carrier mobility can be interpreted by the scattering rates, which can be decomposed by phonon energy.
Here, we focus on the electron mobility in InAs monolayer.
Considering the energy of $3/2k_BT = 39$~meV away from the band edge (see details
  in Ref.~\cite{Ponce2019Feb}), the phonon energy resolved scattering rates $\tau
    ^{-1}$, computed in the presence or in the absence of both corrections, are
  respectively presented in Figs.
\ref{fig:quadfreq}(a) and (b).
The total scattering rate is enhanced by the absence of corrections from 48 to
  68~THz, increased by $\sim40\%$, demonstrating the main influence is caused by
  the dynamical quadrupole.
In both cases, the scatterings are determined by the high-frequency phonons,
  i.e., the LO mode.
Indeed, InAs is a polar material with one isotropic and parabolic
  conduction band, and its behavior can be described by the \Frohlich
  model~\cite{Herbert1952Dec}.
Since quadrupole has negligible influence on the LO mode, the scatterings
  contributed by the LO mode (the area of the right peak) is nearly unchanged.
On the other hand, the quadrupole strongly affects the ZO mode.
When the quadrupole is neglected, the scattering contributed by the ZO mode is
  enhanced from 7.2 to 12.2~THz given by the areas of middle peaks, explaining
  the mobility underestimation by the absence of quadrupole.
The above discussion illustrates that the impacts of Berry connection are materials dependent. 
More importantly, the dynamical quadrupole can strongly affect the computed
  mobility by up to 40\% for materials with high mobilities, illustrating the
  necessity to include quadrupole for their accurate predictions.

\subsection{Impact of spin-orbit coupling}

\begin{figure}[t]
    \includegraphics[width=0.4\paperwidth]{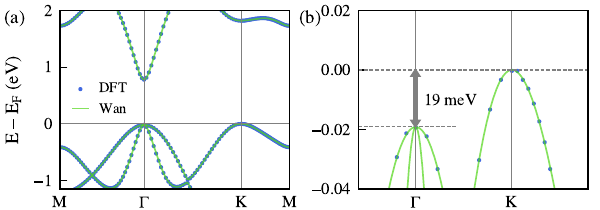}
    \caption{\label{fig:socbands} 
      InAs monolayer electronic band structure (a) calculated
      without SOC, and the corresponding zoomed-in figure (b) around
      Fermi energy.
    }
  \end{figure}

Arising from the interaction between the spin and orbital angular momenta, SOC
  impacts the motion of electrons and modifies the electronic structures,
  especially for materials with heavy atoms.
In the fully-relativistic calculation including SOC [see
      \cref{fig:ppbands}(b)], the conduction band minimum (CBM) presents a Rashba
  splitting at the $\Gamma$ point, analog to a single valley in the BZ.
In InAs monolayer, the VBM is also located at the $\Gamma$ point where the SOC
  splitting occurs, and $\varepsilon_{\rm K} = {\rm E}_{\rm F} - 88~{\rm meV}$.
In the scalar-relativistic case [see \cref{fig:socbands}], the edge of the
  conduction band becomes a single valley at the $\Gamma$ point, and the valence
  bands are changed: the VBM is moved to the K point, the degeneracy at the
  $\Gamma$ point is recovered due to crystal symmetry, and
  $\varepsilon_{\Gamma} = {\rm E}_{\rm F} - 19~{\rm meV}$.
The modifications are even more striking in GaSb and InSb, see
  \cref*{supp-fig:GaSbB} and \cref*{supp-fig:InSbB} in SM~\cite{SI2023}.
In contrast to the electronic structure, SOC induces a negligible change to the
  phonon dispersion as shown in \cref*{supp-fig:InAsB} in SM~\cite{SI2023}, since
  the phonon dispersion is associated with lattice vibrations which are primarily
  determined by the masses of the nuclei and the interatomic forces in the
  lattice.
The deformation potential from the SR calculation is very similar to that from
  the FR calculation, see \cref*{supp-fig:InAsD} in SM~\cite{SI2023}.

  \begin{figure}[b]
  
    \includegraphics[width=0.4\paperwidth]{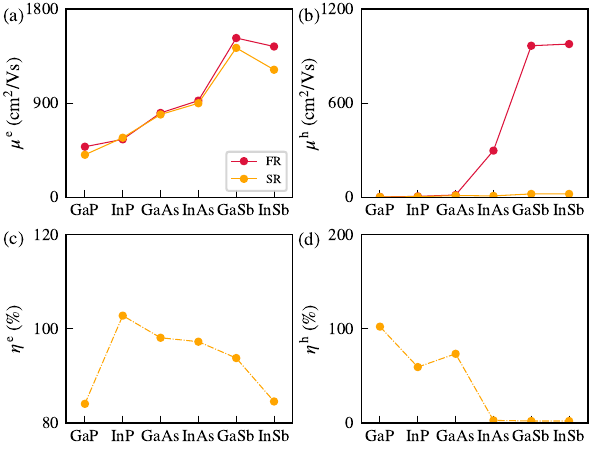}
    \caption{\label{fig:socmob}
    Electron (a) and hole (b) mobilities calculated with FR and SR
  according to \#2 and \#6 methodologies. 
  Taking the FR result as a benchmark,
  variations of the SR result are given as ratios in (c) for electron and
  (d) for hole, respectively. 
    }
  \end{figure}

The comparisons between the FR and SR calculated mobilities are shown in
  \cref{fig:socmob}.
The absence of SOC causes limited changes in electron mobilities for all
  materials, while more prominent variations emerge in hole mobilities.
For phosphides and GaAs, the hole mobilities are strongly suppressed in both
  cases, since the VBMs are always located at the K points in these materials.
For InAs and antimonides, the hole mobilities are greatly suppressed by the
  absence of SOC.

\begin{figure}[t]

  \includegraphics[width=0.3\paperwidth]{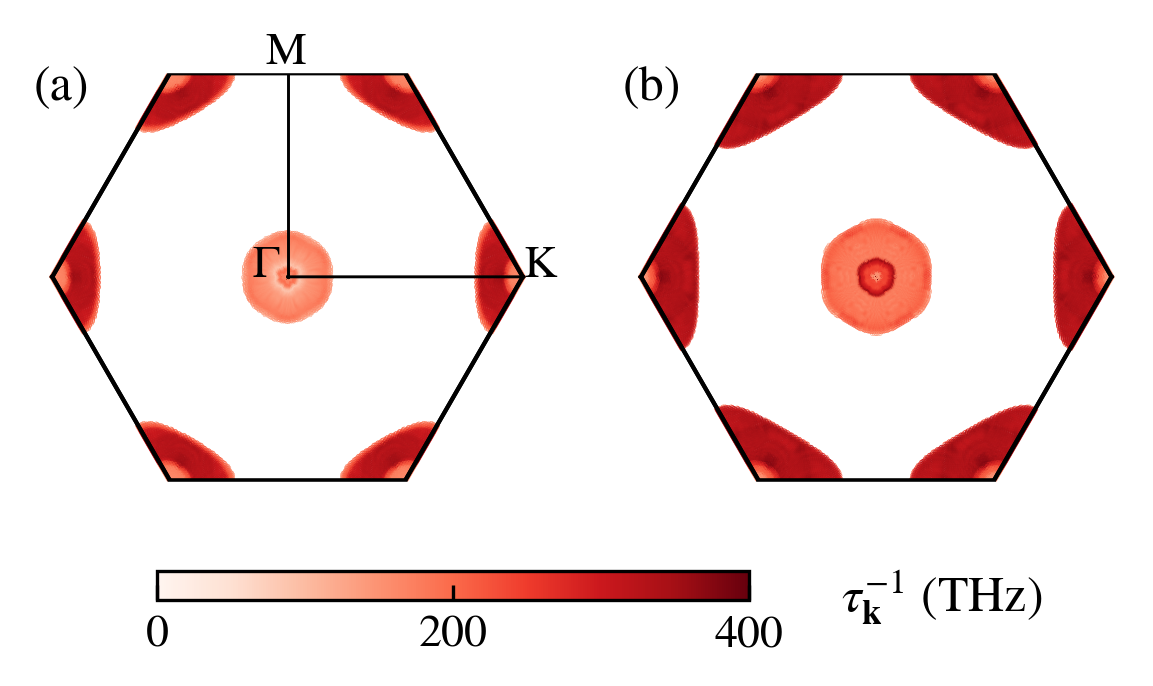}
  \caption{\label{fig:soc_inv_tau}
    Momentum-resolved scattering rates of holes in InAs monolayer in (a)~FR and
      (b)~SR calculations, given by $\tau^{-1}_{\bf k}~=~\frac{1}{N^{\rm
              w}}\sum_{n}\tau^{-1}_{n {\bf k}}$.
  }
\end{figure}

\begin{figure}[b]
  \includegraphics[width=0.34\paperwidth]{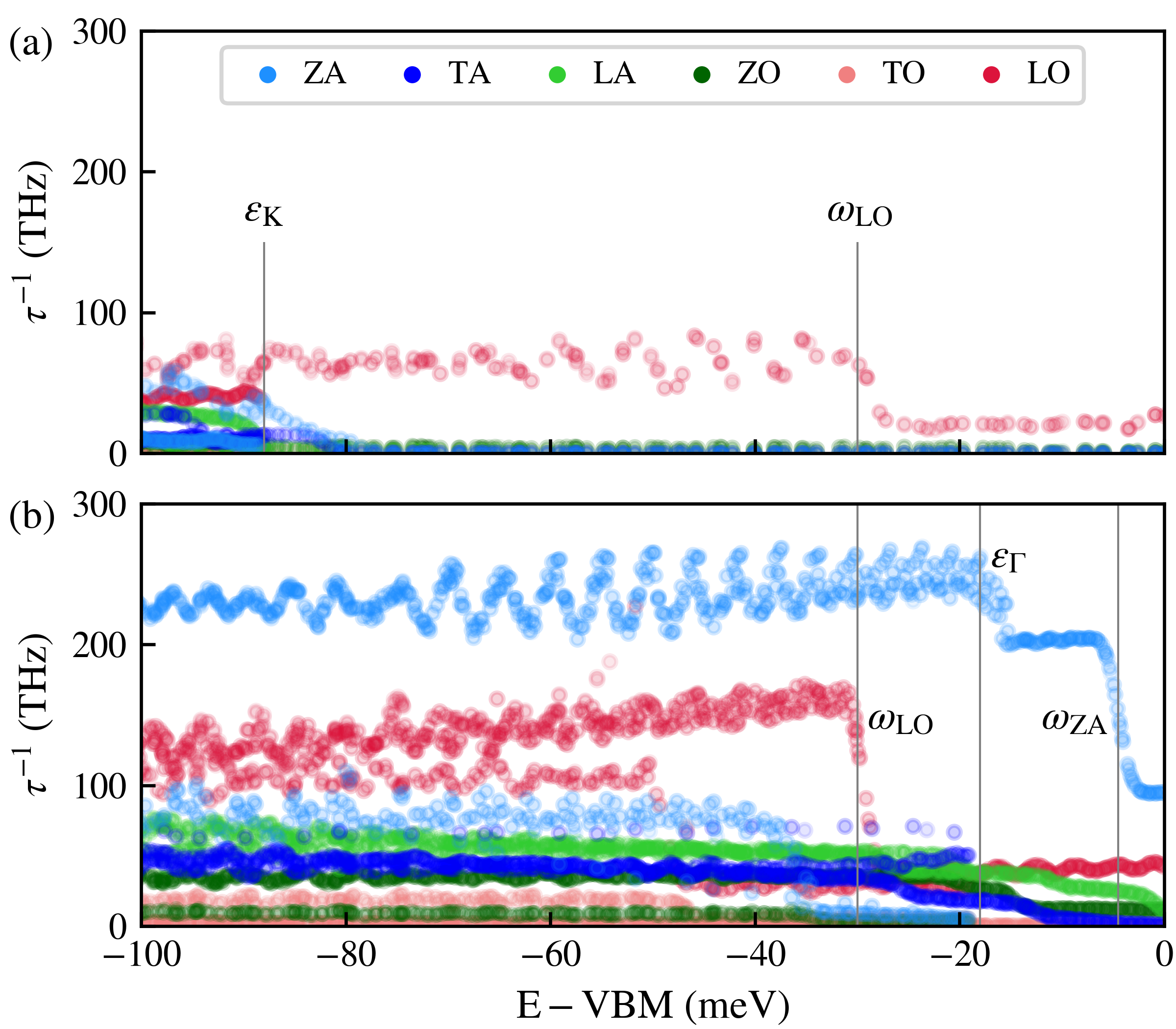}
  \caption{\label{fig:mode_inv_tau}
    Mode-resolved scattering rates of hole mobility in InAs monolayer in (a)~FR and (b)~SR
      calculations.
    % Note $\omega_{\rm ZA}\approx4$~meV at the ${\bf q}={\rm K}$ point.
  }
\end{figure}

The significant influence caused by SOC can be explained by the scattering
  rates of holes.
The $\bf k$-resolved scattering rates in FR and SR calculations of InAs
  monolayer are respectively presented in Figs.~\ref{fig:soc_inv_tau}(a) and (b).
Compared with the FR scattering rates, SR scattering rates are enhanced
  especially around the K points.
This is because in the FR calculation, the VBM is located at the $\Gamma$ point
  which is a single scattering peak.
Besides, eigenstates at the $\Gamma$ and K points are separated by a large
  energy of 88~meV, increasing the barrier of scattering between $\Gamma$ and K.
In contrast, in the SR case, the VBMs are located at the K points which present
  multi-peaks of scattering in the BZ.
Moreover, the energy barrier between eigenstates at the $\Gamma$ and K points
  is reduced to 19~meV, increasing the interpeak scattering.
Consequently, the hole mobility in the SR calculation is notably reduced.

Additional understanding of the hole mobility can be gained by mode-resolved
  scattering analysis, whose behavior is determined by the electron and phonon dispersions
  due to the energy and momentum conservations.
\Cref{fig:mode_inv_tau}(a)
shows that in the FR calculation,
the LO mode makes the dominant contribution
in the energy range from VBM to  $\varepsilon_{\rm K} = -88 $~meV,
since  there is only one band peak at the $\Gamma$ point,
analog  to a \Frohlich system.
The increase at $-30$~meV is attributed to the LO mode with $\omega_{\rm
      LO}\approx30$~meV.
Below $\varepsilon_{\rm K}$, multi-peaks at K points are involved in the Fermi
  surface window, and the ZA mode starts to contribute to the intrapeak and
  interpeak scatterings with small phonon energies. 
As indicated by \cref{fig:mode_inv_tau}(b), in the FR case, ZA phonons dominate
  the scattering due to the intrapeak scattering at K point.
A sharp increase of ZA contribution occurs around $-4$~meV since more phonons
  of $\omega_{\rm ZA}\approx4$~meV at ${\bf q}={\rm K}$ point [see \cref{fig:ppbands}(e)] are enabled for
  interpeak scatterings between K points in ${\bf k}$-space.
When $\varepsilon_{\Gamma} = -19$~meV enters into the Fermi surface window, ZA
  contribution is further enhanced by the interpeak scattering between the
  $\Gamma$ and K points.
Furthermore, expanding the Fermi surface window allows LO mode with
  $\omega_{\rm LO}\approx30$~meV to participate in the scattering.
Since more phonons are involved, the scattering is enhanced in the SR
  calculation.
The above discussion shows that SOC can significantly modify the electronic
  structure and lead to large mobility variations $\sim100\%$, demonstrating the
  importance of considering SOC, especially for materials with heavy atoms and
  multi-band edges.

\subsection{Balance between accuracy and cost}
\begin{figure}[b]
  \includegraphics[width=0.4\paperwidth]{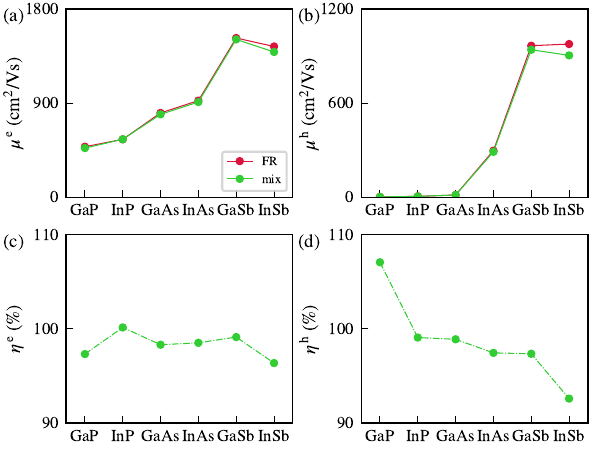}
  \caption{\label{fig:mixmob}
  Electron (a) and hole (b) mobilities calculated with the FR
  and the mix scheme according to \#2 and \#7 methodologies. 
  Taking the FR
  result as a benchmark, variations of the mix methodology are given as
  ratios in (c) for electron and (d) for hole, respectively.  
  }
\end{figure}
Although SOC induces significant changes in the electronic structures, it
  negligibly affects the phonon dispersion and scattering potential variation.
Considering the cost of DFPT calculations, we can use the electronic
  wavefunctions from the FR calculation and the scattering potential variations
  from the SR calculation to perform the Wannier interpolation for EPC
  computation.
The quality of the interpolation has been validated by the corresponding Wannier
  and DFPT deformation potentials as shown in \cref*{supp-fig:InAsD} in
  SM~\cite{SI2023}. 

\Cref{fig:mixmob}
presents
the comparison between the complete FR calculation
and the mix methodology.
For the electron mobility, the discrepancies are limited below 3\%, while for
  the hole mobility, the discrepancies can be up to 7\% in GaP and InSb.
Still, the mix methodology produces a result consistent with the complete FR
  calculation, demonstrating its effectiveness.

\begin{figure}[t]

  \includegraphics[width=0.38\paperwidth]{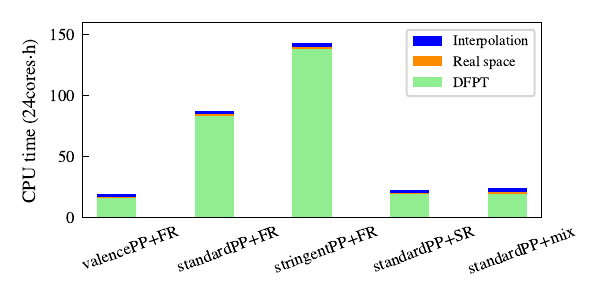}
  \caption{\label{fig:cost}
    Computational costs for different methodologies in InAs monolayer.
    DFPT denotes the phonon calculation for potential variation with a
      self-consistency threshold of $10^{-18}$.
    Real space denotes the EPW calculation for representation transform from
      reciprocal to real space on the coarse grid.
    Interpolation denotes the interpolation from real to reciprocal space for
      transport properties on fine grid.
    All the calculations are performed using 24 AMD EPYC 7H12 cores with a base
      clock speed of 2.6~GHz.
  } \end{figure}

Although paying a small accuracy loss, the mix methodology can significantly
  enhance the calculation efficiency.
The computational costs of various methodologies mentioned above are summarized
  in \cref{fig:cost} where dynamical quadrupole and Berry connection are always
  considered.
DFPT is always the most expensive calculation for all methodologies.
The valencePP+FR and standardPP+SR methodologies are inexpensive but
  unreliable.
Both standardPP+FR and stringentPP+FR produce consistent results, while the
  latter is more expensive since more semi-core states are included.
The standardPP+mix methodology exhibits the same computational cost as the
  standardPP+SR in DFPT calculation, and the similar computational cost to the
  standardPP+FR in EPW calculation.
Consequently, the total cost of standardPP+mix methodology significantly
  decreases, which is 28\% of the standardPP+FR cost, and 17\% of the
  stringentPP+FR cost.
The above discussion suggests that the standardPP+mix methodology can strike a
  good balance between accuracy and cost.

\section{Temperature-dependent mobilities}

\begin{figure}[tb]
  \includegraphics[width=0.4\paperwidth]{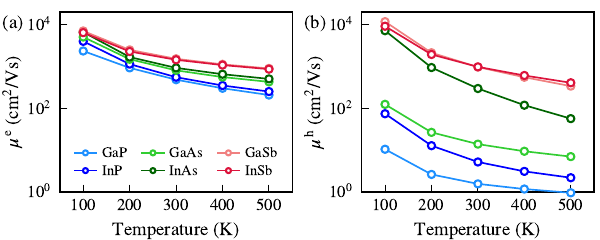}
  \caption{\label{fig:mobtemp}
    Temperature-dependent (a)~electron and (b)~hole  drift mobilities.
  }
\end{figure}

To show the most accurate results, we present the temperature-dependent
  mobilities calculated by the standardPP+Quad+Berry+FR methodology in
  \cref{fig:mobtemp}.
All the mobilities go down with increasing temperature.
For the electron mobilities, all materials present a high \mue over 100~\unitu,
  attributed to the single valley of conduction bands.
The mobilities increase with the atomic number due to the decreasing effective
  mass, and the antimonides can reach \mue$\approx10^4$~\unitu at 100~K.
The decay of \mue can be one order of magnitude from 100~K to 500~K.
For the hole mobilities, \muh decays much faster with temperature.
Still, antimonides can present high hole mobilities thanks to their
  single-peak VBMs.

Instead of drift mobility, Hall mobility is more commonly measured experimentally, since it can be determined using a well-established setup, i.e.,
  the Hall effect measurement.
Indeed, such a transport measurement is performed under an external magnetic field, which
  induces a Lorentz force on the carriers and then changes the drift mobility
  $\mu$ to be the Hall mobility $\mu^{\rm H}$.
Denoting the ratio between the Hall mobility and the drift mobility, the Hall
  factor is commonly assumed to be unity~\cite{Schroder2015Jun}.
However, the suitability of this assumption should be carefully evaluated.
The Hall mobility can be calculated by adding the external magnetic field to
  the BTE, see details in Eqs.~(6)-(9) in Ref.~\cite{Lee2023Feb}.
In the present computations of Hall factors, we apply an external magnetic field along
  $z$ direction with ${B_z} = 10^{-10}$~T, and we focus on the tensor element
  $r^{\mathrm{H}}_{xy}$ which indicates the ratio of $\mu^{\mathrm{H}}_{xy}$ over
  $\mu_{xx}$, the subscripts will be omitted in the following discussions.

\begin{figure}[tb]
  \includegraphics[width=0.4\paperwidth]{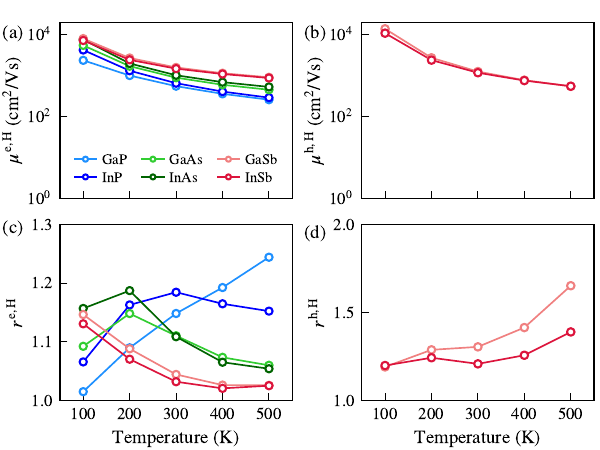}
  \caption{\label{fig:halltemp}
    Temperature-dependent
    (a)~electron and (b)~hole Hall mobilities, 
    as well as 
    (c)~electron and (d)~hole Hall factors.
  }
\end{figure}

\Cref{fig:halltemp} presents
the temperature-dependent Hall mobilities  and Hall factors,
depicting the materials with high mobilities which are more promising for
  experiments.
For the electron transport, the magnetic field introduces limited changes to
  the mobility, as demonstrated by \reH which is below 1.3 in all the materials.
However, for the hole transport, \rhH can be as large as 1.7, as shown by the GaSb
  monolayer at high temperature.
These theoretical results demonstrate that the unitary Hall factor can be a rough
  approximation, which may sometimes lead to the overestimation of the drift mobility in
  experiments.

\section{Summary}
Carrier mobilities investigated using DFPT and Wannier functions
are reported herewith for 6  semiconductors in the III-V monolayer family.
The quality of the Wannier interpolation has been validated by the comparison
  with DFPT calculations.
To show the impacts induced by different approximations, several methodologies
  have been proposed to investigate the influences on mobilities caused by
  semi-core states, dynamical quadrupole, Berry connection, and spin-orbit
  coupling.
Semi-core states in pseudopotentials are found to be essential for
  accurate mobility calculations, dynamical quadrupole can induce a variation of
  $\sim40$\%, Berry connection causes an impact of $\sim10$\%, and SOC can even
  yield an influence of $\sim100$\% for materials with multi-peak electronic
  structures.
The different mechanisms are interpreted by the momentum- and mode-resolved scattering
  rates.
Besides, DFPT results are negligibly affected by SOC, which can be neglected to
  accelerate computation with a precision loss of less than 7\%.
After evaluating the various methodologies, temperature-dependent
  drift mobilities are computed, illustrating that the temperature can change the mobilities
  by one or two orders of magnitude, and the Hall factors range from 1.0 to 1.7.
From the point of view of $ab~initio$ modeling, it is mandatory and commonly accepted to
  implement certain approximations.
Consequently, it is crucial to understand the extent of their impacts and
  evaluate their suitability.
Striking the balance between accuracy and cost, this research can definitely provide
  guidelines for accurate and efficient calculations of carrier mobilities in 2D
  semiconductors.

\begin{acknowledgments}
The authors would like to thank Matteo Giantomassi for fruitful discussions.
S.~P. 
acknowledges the support from the Fonds de la Recherche Scientifique de Belgique (\frs-FNRS).
J.~Z. and J.-C.C. acknowledge financial support from the F\'ed\'eration Wallonie-Bruxelles through the ARC Grant ``DREAMS'' (No. 21/26-116), from the EOS project ``CONNECT'' (No. 40007563), and from the Belgium \frs-FNRS through the research project (No.~T.029.22F).
Computational resources have been provided by the PRACE award granting access
  to MareNostrum4 at Barcelona Supercomputing Center (BSC), Spain and Discoverer
  in SofiaTech, Bulgaria (OptoSpin project ID. 2020225411), and by the Consortium des Équipements de Calcul Intensif (C\'ECI), funded by the \frs-FNRS under Grant No. 2.5020.11 and by the Walloon Region, as well as computational resources awarded on the Belgian share of the EuroHPC LUMI supercomputer.
\end{acknowledgments}

% \bibliography{ref.bib}
\input{bbl.bbl}
\end{document}